\documentclass[traditabstract]{aa} 
\usepackage{graphicx}
\usepackage{txfonts}
\usepackage{natbib}
\usepackage{color}
\begin{document}

   \title{The subgiant branch of $\omega$~Cen seen through high-resolution
          spectroscopy. I. The first stellar generation in 
          $\omega$~Centauri?\thanks{Based on data collected at the ESO  VLT in
          Chile, with UVES and FLAMES under programs 68.D-0332(A) and
          079.D-0021. Also based on literature data from the ESO WFI, under
          programs 62.L-0354 and 63.L-0439, and on data  collected from the
          following online databases: NIST, VALD, Kurucz, GEISA.}}

\titlerunning{The first stellar generation in $\omega$~Centauri} 

   \author{E. Pancino \inst{1},  A. Mucciarelli\inst{2}, 
           L. Sbordone\inst{3,4}, M. Bellazzini\inst{1}, L. Pasquini\inst{5}, 
	   L. Monaco\inst{6}, F. R. Ferraro\inst{2}}
   \authorrunning{Pancino et al.}
   \institute{INAF-Osservatorio Astronomico di Bologna, via Ranzani 1, I-40127,
	     Bologna, Italy\\
             \email{elena.pancino@oabo.inaf.it}
             \and
	Dipartimento di Astronomia, Universit\`a degli studi di Bologna, via
	     Ranzani 1, I-40127, Bologna, Italy 
	     \and
	Max Planck Institute for Astrophysics, Karl-Schwarzschild-Str. 1, D-85741
             Garching, Germany  
	     \and
	Laboratoire d'Etudes des Galaxies, Etoiles, Physique et Instrumentation
             GEPI, Observatoire de Paris, Avenue de L'Observatoire 61, F-75014, 
	     France
	     \and     
	European Southern Observatory, Karl-Schwarzschild-Strasse 2, D-85748 
	     Garching bei M\"unchen, Germany    
             \and
        European Southern Observatory, Casilla 19100, Santiago, Chile         
             }

   \date{Received September XX, XXXX; accepted March XX, XXXX}

 
   \abstract{We analysed high-resolution UVES spectra of six stars belonging to
   the subgiant branch of $\omega$~Centauri, and derived abundance ratios of 19
   chemical elements (namely Al, Ba, C, Ca, Co, Cr, Cu, Fe, La, Mg, Mn, N, Na,
   Ni, Sc, Si, Sr, Ti, and Y). A comparison with previous abundance
   determinations for red giants provided remarkable agreement and allowed us to
   identify the sub-populations to which our targets belong. We found that three
   targets belong to a low-metallicity population at [Fe/H]$\simeq$--2.0~dex,
   [$\alpha$/Fe]$\simeq$+0.4~dex and [s/Fe]$\simeq$0~dex. Stars with similar
   characteristics were found in small amounts by past surveys of red giants. We
   discuss the possibility that they belong to a separate sub-population that we
   name VMP (very metal-poor, at most 5\% of the total cluster population),
   which -- in the self-enrichment hypothesis -- is the best-candidate first
   stellar generation in $\omega$~Cen. Two of the remaining targets belong to
   the dominant metal-poor population (MP) at [Fe/H]$\simeq$--1.7~dex, and the
   last one to the metal-intermediate (MInt) one at [Fe/H]$\simeq$--1.2~dex. The
   existence of the newly defined VMP population could help to understand some
   puzzling results based on low-resolution spectroscopy (Sollima et al.,
   Villanova et al.) in their age differences determinations, because the
   metallicity resolution of these studies was probably not enough to detect the
   VMP population. The VMP could also correspond to some of the additional
   substructures of the subgiant-branch region found in the latest HST
   photometry (Bellini et al.). After trying to correlate chemical abundances
   with substructures in the subgiant branch of $\omega$~Cen, we found that the
   age difference between the VMP and MP populations should be small
   (0$\pm$2~Gyr), while the difference between the MP and MInt populations could
   be slightly larger (2$\pm$2~Gyr). }


   \maketitle
%

\section{Introduction}
\label{sec-intro}

Much has been written on $\omega$~Cen, which is probably the most studied
cluster in the Milky Way \citep{fvl02}. From the pioneering studies in the 60s
to the latest high-quality data and models, more and more details of its
multiple stellar populations have come to light, but the picture did not become
as clear as expected. Excellent photometries and astrometric catalogues have
recently been produced by both ground-based \citep[such
as][]{lee99,p00,fvl00,hilker00,hughes00,sollima05a,calamida09,bellini09} and
space telescopes \citep[e.g.,][]{ferraro04,bedin04,bellini10}, complemented by
high-quality spectroscopic abundance studies both at high
\citep[e.g.,][]{norris95,smith00,cunha02,p03,johnson08,johnson09,johnson10,marino10}
and low \citep{norris96,suntzeff96,sollima05b,stanford06,villanova07}
resolution. Nevertheless, several puzzles still await solutions.

One of the open problems concerns the most sensitive region of the colour
magnitude diagram (hereafter CMD) to age differences: the subgiant branch (SGB).
A large number of photometric and low-resolution spectroscopic studies
\citep{hughes00,hilker00,p03,hughes04,hilker04,rey04,ferraro04,stanford06} found
age spreads ranging from 2 to 6~Gyr, with a few exceptions and puzzles (see
Section~\ref{sec_ages}, for more details). One example of the difficulties
encountered in the study of this complex region is posed by the two studies by
\citet{sollima05b} and \citet{villanova07}, who used the same ACS dataset and
low-resolution spectra of similar quality, but reached opposite conclusions on
the total age spread -- and age distribution -- of $\omega$~Cen. Still, even
with the best ACS photometries \citep[see, e.g.][]{bellini10}, it is not easy to
understand which features of the SGB region correspond to each of the
populations that are spectroscopically identified on the red giant branch (RGB),
which are known in great detail thanks to the above cited works. This
understanding is crucial to solve the relative ages problem in $\omega$~Cen, and
to derive the age-metallicity relation, a fundamental ingredient of any model
for the formation and evolution of this unique stellar system.

\begin{table*}
\caption{Observing Logs.}
\label{tab_logs}
\begin{center}
\begin{tabular}{l c c c c c c c c c l} 
\hline\hline
\noalign{\smallskip}
ID$_{WFI}$ & R.A. (J2000) & Dec (J2000) & V & (B-V)$_0$ & (V-I)$_0$ & R & t$_{exp}^{(tot)}$ & S/N & $V_r$ & Notes \\
& (deg) & (deg) & (mag) & (mag) & (mag) & ($\lambda/ \delta\lambda$) & (sec) & (@550nm) & (km s$^{-1}$) & \\
\noalign{\smallskip}
\hline
\noalign{\smallskip}
503358 & 201.960086 & -47.759426 & 17.53 & 0.65 & 0.65 & 45\,000 & 10800 & 50 & 231.0 $\pm$ 0.2 & \rm{Lower SGB} \\
503951 & 201.870201 & -47.746107 & 17.58 & 0.65 & 0.65 & 45\,000 & 11400 & 50 & 235.8 $\pm$ 0.1 & \rm{Lower SGB} \\
507109 & 201.848173 & -47.692322 & 17.31 & 0.60 & 0.66 & 45\,000 & 11288 & 50 & 229.7 $\pm$ 0.5 & \rm{Upper SGB} \\
507633 & 201.813303 & -47.685166 & 17.39 & 0.56 & 0.66 & 45\,000 & 12000 & 45 & 226.7 $\pm$ 0.3 & \rm{Upper SGB} \\
512115 & 201.846338 & -47.637524 & 17.64 & 0.65 & 0.66 & 45\,000 & 11400 & 45 & 241.4 $\pm$ 0.1 & \rm{Lower SGB} \\ 
512938 & 201.804070 & -47.630694 & 17.32 & 0.60 & 0.65 & 45\,000 & 10000 & 50 & 234.6 $\pm$ 0.2 & \rm{Upper SGB} \\
\hline\hline	  
\end{tabular}
\end{center}
\end{table*}

In this paper we present the analysis of a set of UVES high-resolution spectra of
six SGB stars, selected from the wide field photometric catalogue by \citet{p00}
and \citet{p03}. A preliminary analysis of the same dataset was presented by
\citet{p03}. We describe the spectra reductions in Section~\ref{sec-data}; the
abundance analysis in Section~\ref{sec-abo}; and the abundance ratio results in
Section~\ref{sec-res}. Our main results are discussed in detail in
Section~\ref{sec-disc} and are summarized in Section~\ref{sec-concl}.

\begin{figure}
\centering
\includegraphics[width=\columnwidth]{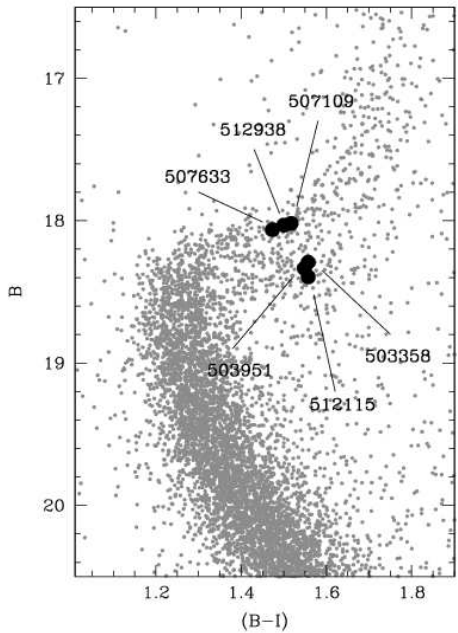}
\caption{Location of the programme stars on the CMD of $\omega$~Cen. The WFI B,
B--I photometry of stars in CCD\#5 \citep[from][]{p00} is shown as grey dots.
The six UVES targets are marked (black filled circles) with their WFI  catalogue
numbers.}
\label{fig_cmds}
\end{figure}

\begin{figure}
\centering
\includegraphics[width=\columnwidth]{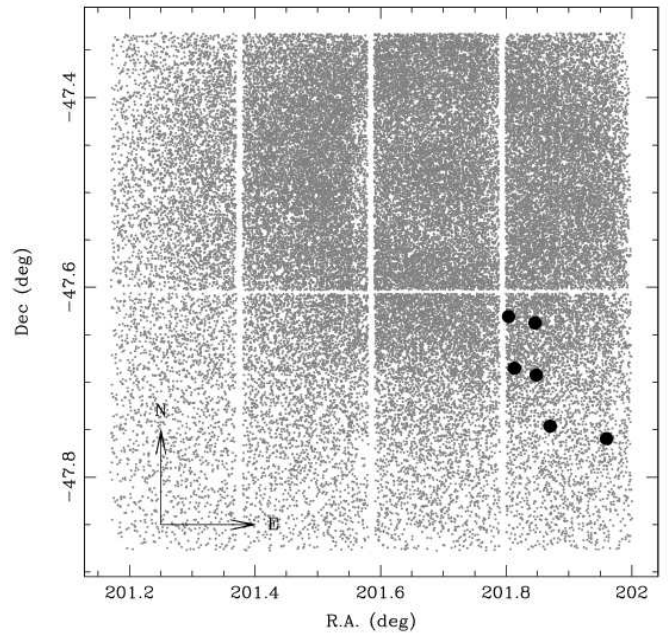}
\caption{Location of the programme stars on the area of $\omega$~Cen. Grey dots
mark stars belonging to the WFI photometry by \citet{p00}. Filled circles mark
the postion of the six UVES targets.}
\label{fig_area}
\end{figure}

\section{Observations and data reduction}
\label{sec-data}

We selected our six targets from the WFI B and I photometry presented in
\citet{p00}, complemented with V magnitudes from \citet{p03}. The coordinates
were obtained using the astrometric catalogue by \citet{fvl00}. As shown in
Figures~\ref{fig_cmds} and \ref{fig_area}, the SGB region of $\omega$~Cen in the
external parts of the cluster shows clear substructures (all our targets lie on
the WFI CCD\#5). This is more clearly seen in other literature photometries
\citep{ferraro04,bedin04,bellini10}, but they are obtained from space, with the
HST, in the very centre of $\omega$~Cen, where UVES follow-up from the ground
would be difficult. Three of the programme stars were selected towards the upper
envelope of the SGB and another three towards the lower envelope.

Echelle spectroscopy was obtained on 18--20 March 2002, with UVES at the ESO
Kueyen (VLT~UT2) telescope, on Cerro Paranal, Chile. Table~\ref{tab_logs}
reports the log of the observations, along with some basic target information.
Each star was observed twice and on different nights, both to minimize the
cosmic rays impact and to identify possible radial velocity shifts. Given the
faint magnitude of these stars (V$\simeq$17.5), the spectra were binned on chip
(2$\times$1), so that the resolution element is covered by $\simeq$3 pixels. A
final S/N$\simeq$50 per pixel was achieved around 550~nm.

\begin{table*}
\caption{Equivalent widths and atomic data for the lines used in the classical
abundance analysis of the program stars. The complete version of the table is
available at CDS. Here we show a few lines to illustrate its
contents.}             
\label{tab-ew}      
\centering          
\begin{tabular}{r r r r r r r r r r r r r r r r r}     
\hline\hline 
& & & & \multicolumn{3}{c}{Star WFI~503358} & \multicolumn{3}{c}{Star WFI~503951} & & \multicolumn{3}{c}{Star WFI~512115} & \multicolumn{3}{c}{Star WFI~512938} \\      
$\lambda$ & El & $\chi_{\rm{ex}}$ & $\log gf$ & EW & $\delta$EW & Q & EW & $\delta$EW & Q & ... & EW & $\delta$EW & Q & EW & $\delta$EW & Q \\
(\AA) & & (eV) & (dex) & (m\AA) & (m\AA) & & (m\AA) & (m\AA) & & & (m\AA) & (m\AA) & & (m\AA) & (m\AA) \\
\hline     	     
4283.01 & CaI & 1.89 & -0.29 & 77.0 & 8.9 & 1.463 & 95.5 & 12.2 & 1.832 & ... &  0.0 &  0.0 & 0.000 & 67.0 & 5.6 & 1.248 \\
4289.37 & CaI & 1.88 & -0.39 & 75.6 & 8.0 & 1.566 & 80.2 &  7.3 & 1.378 & ... & 79.7 & 12.7 & 1.492 & 64.5 & 6.5 & 1.213 \\
4298.99 & CaI & 1.89 & -0.51 &  0.0 & 0.0 & 0.000 &  0.0 &  0.0 & 0.000 & ... &  0.0 &  0.0 & 0.000 &  0.0 & 0.0 & 0.000 \\  
4302.53 & CaI & 1.90 &  0.18 &  0.0 & 0.0 & 0.000 &  0.0 &  0.0 & 0.000 & ... &  0.0 &  0.0 & 0.000 &  0.0 & 0.0 & 0.000 \\ 
4318.65 & CaI & 1.90 & -0.29 & 72.8 & 8.1 & 1.876 & 93.6 &  7.5 & 1.747 & ... &  0.0 &  0.0 & 0.000 & 61.1 & 3.7 & 0.851 \\
4434.96 & CaI & 1.89 &  0.07 &  0.0 & 0.0 & 0.000 &  0.0 &  0.0 & 0.000 & ... &  0.0 &  0.0 & 0.000 & 81.0 & 5.9 & 1.227 \\
\hline \hline                 
\end{tabular}
\end{table*}

The red spectra (upper and lower red CCDs) were reduced with the ESO-UVES
pipeline \citep{uves}, which semi-automatically performs bias correction,
flat-field correction, inter-order background subtraction, optimal extraction
with cosmic-ray rejection, wavelength calibration (with rebinning), and final
merging of all overlapping orders. However, since the S/N ratio is significantly
lower on the blue part of the spectra (S/N$\simeq$25 around 450~nm), we decided
to manually perform the echelle reduction for the blue CCD with the {\it
noao.imred.ccdred} and {\it noao.imred.echelle} packages within
IRAF\footnote{http://iraf.noao.edu/. IRAF is distributed by the National Optical
Astonomy Ob\-ser\-va\-to\-ries, which is operated by the association of
Universities for Research in Astronomy, Inc., under contract with the National
Science Foundation.}.

The two one-dimensional and wavelength-calibrated spectra obtained for each star
were normalized by fitting their continua with a cubic spline, then corrected
for the main telluric absorption features ({\it noao.onedspec.telluric} in
IRAF), using as a reference a hot, fast rotating star (HR~5206/HD~120640)
selected from the {\it Bright Star Catalog} \citep{bsc}, and observed each night
at an airmass not too different from the targets. Finally the spectra were
summed -- after correcting for radial velocity shifts -- to produce one single
spectrum for each of the six observed stars.

Radial velocities were obtained with
DAOSPEC\footnote{http://www3.cadc-ccda.hia-iha.nrc-cnrc.gc.ca/community/STETS
ON/daospec/ ;  http://www.bo.astro.it/$\sim$pancino/projects/daospec.html}
\citep{daospec} and the procedure described by \citet{p10}. In short, the
laboratory wavelength of selected absorption lines (Section~\ref{sec_ew}) was
used to measure the observed radial velocity. The heliocentric correction was
computed with IRAF and the telluric H$_2$O and O$_2$ absorption bands redward of
580~nm were used to correct for zeropoint shifts. All six stars were radial
velocity members of $\omega$~\,Cen, considering an average V$_r$=232.8 or 233.4
km s$^{-1}$, as determined by \citet{meylan95} and \citet{p07}, respectively,
with a central velocity dispersion of the order of 20 km s$^{-1}$. The resulting
velocities and errors are listed in Table~\ref{tab_logs}.

\section{Abundance analysis}
\label{sec-abo}

\subsection{Equivalent widths and atomic data}
\label{sec_ew}

We selected the majority of our lines and of their atomic data from the
VALD\footnote{http://www.astro.uu.se/$\sim$vald/} database \citep{vald}. To
identify reliable lines, high S/N median spectra of the six UVES targets were
created and the cleanest, unblended lines of the available elements were
identified. Only lines that appeared in at least three of the six stars were
retained in our preliminary selection. DAOSPEC was used to measure the
equivalenth widths (EW) of all the chosen lines. A first-pass abundance analysis
was performed (Section~\ref{sec_abo}): lines that showed systematically higher
errors and bad Q parameters \citep[see][for details]{daospec,p10} and that
simultaneously gave systematically discrepant abundances were rejected. Finally,
all lines that gave EW$<$15~m\AA, or EW$>$100~m\AA\  (with a few exceptions, see
Section~\ref{sec_abo}) were not used to determine abundances. The DAOSPEC EW
measurements used for the abundance analysis are shown in Table~\ref{tab-ew},
along with the formal error $\delta$EW and the quality parameter Q for each line
\citep{daospec}.

\begin{figure}
\centering
\includegraphics[bb=10 380 590 720,clip,width=\columnwidth]{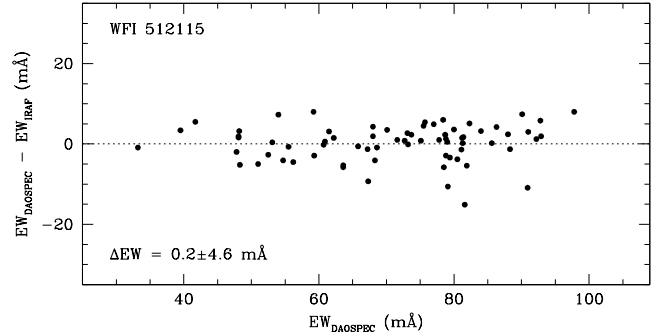}
\caption{Comparison between DAOSPEC and IRAF {\em splot} measurements for star
WFI~512115.}
\label{fig_ew}
\end{figure}

A few lines were measured with the help of spectral synthesis because they were
reported to have significant hyperfine structure (HFS). In particular, we used 
the atomic data by \citet{martin88} for the Mn\,I lines at 4030, 4033, 4034, 
4041, and 4055~\AA; the
NIST\footnote{http://physics.nist.gov/PhysRefData/ASD/index.html} atomic data
for the Ba\,II lines at 4934, 5853, 6141, and 6496~\AA; the atomic data by
\citet{lawler01} for the La\,II lines at 3988, 4086, 4123, and 4238~\AA; and the
atomic data by \citet{bielski75} for the 5105~\AA\  Cu\,I line\footnote{We could
not use the 5872~\AA\  Cu\,I line since it falls into the gap between the two
UVES red CCDs.}. For the CH and CN molecular bands, we used the Kurucz molecular
linelists\footnote{http://kurucz.harvard.edu/LINELISTS/LINESMOL/}, but we had
to  revise the log$gf$ values of C downwards by 0.3~dex, similarly to what was
reported by \citet{bonifacio98}, \citet{lucatello03}, and \citet{spite05} (see
also Section~\ref{sec_abo}). 

\begin{figure}
\centering
\includegraphics[width=\columnwidth]{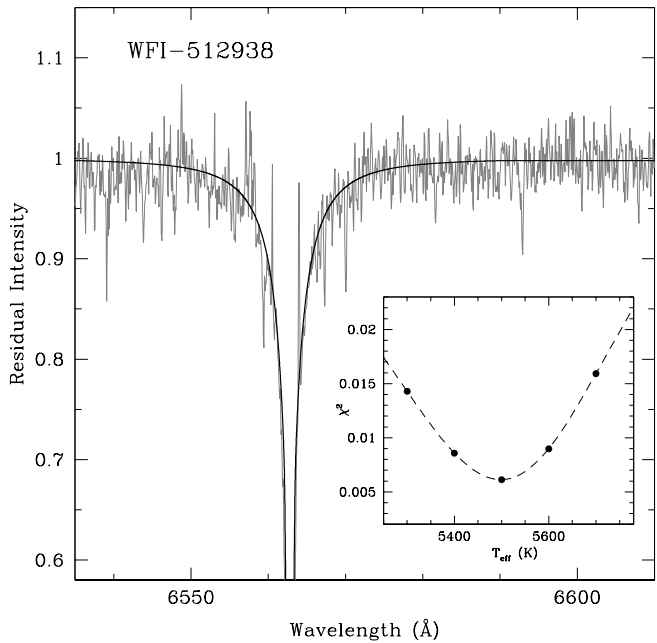}
\caption{Example of the H$_{\alpha}$ profile fit for star WFI~512938. The solid
line represents the best-fitting synthetic spectrum where all parameters are
kept fixed as in Table~\ref{tab_params} except for the effective temperature,
T$_{eff}$. The inset panel shows the run of the $\chi ^2$ of fits with different
values of T$_{eff}$.}
\label{fig_Halpha}
\end{figure}

\subsection{Atmospheric parameters and best-model search}
\label{sec-param}

A first guess of the atmospheric parameters was derived from the WFI photometry.
Dereddened (B--V)$_0$ and (V--I)$_0$ colours were obtained from B, V, and I
magnitudes adopting E(B--V)=0.11 \citep{lub02} and E(V--I)/E(B--V)=1.30
\citep{dean78}, and are listed in Table~\ref{tab_logs} along with the V
magnitudes. The V--I colour was converted from the original V--I$_C$, based on the
Cousins I magnitude, to the V--I$_J$, based on the Johnsons I magnitude, with the
relations by \citet{bessell79}. Effective temperatures (hereafter T$_{eff}$) and
bolometric corrections (BC$_V$) were obtained with the \citet{alonso99}
calibration. Surface gravities (hereafter log\,$g$) were then derived by means of
fundamental relations:

\begin{equation}
\label{eq_params}
\log g_*=0.4(M_V+BC_V)+4\log T_{eff,*}-12.61,
\end{equation}

\noindent where the solar values where assumed in conformity with the IAU
recommendations \citep{iau}, i.e., $\log g_{\odot}=4.437$,
T$_{eff,\odot}$=5770~K and M$_{bol,\odot}$=4.75. A typical mass of
0.8~M$_{\odot}$ was assumed for the programme stars \citep{vdb01}, and we used
(m--M)$_V$=14.04$\pm$0.11~mag \citep{bellazzini04}. An independent estimate of
T$_{eff}$, which is the most influential parameter when determining abundances,
was also derived from the profile fitting of the H$_{\alpha}$ line wings. We
computed Kurucz atmosphere models with ATLAS9 \citep{kurucz93,kurucz05}, using
gravities and global metallicities close to the photometric parameters of each
target. We built synthetic spectra with a modified version of
SYNTHE\footnote{The broadening theory adopted in this routine is that of
\citet{ali65} and \citet{vidal73}.}, exported for the Linux OS
\citep{sbordone04}. An example of a typical H$_{\alpha}$ profile fit is shown in
Figure~\ref{fig_Halpha}. 

We found general agreement between the photometric and
the H$_{\alpha}$ T$_{eff}$ estimates, but the scatter was large, with differences
of up to 300~K in some cases. When averaging estimates from B--V and V--I
together, the scatter went down (see Table~\ref{tab_params}). We decided to rely
on the H$_{\alpha}$ temperatures as a first guess for the spectroscopic analysis.
First guess log $g$ values were derived from H$_{\alpha}$ temperatures using V
magnitudes and the \citet{alonso99} calibration. 

The final atmospheric parameters were then derived through the usual
``spectroscopic method" (see Section~\ref{sec_abo} for abundance calculation
details):  after we had a good number of Fe~I and Fe~II lines (approximately 200
and 10, respectively, depending on the star), we re-adjusted the parameters by
choosing those that minimized {\em (i)} the slope of [Fe~I/H] versus the
excitation potential, $\chi_{ex}$; {\em (ii)} the slope of [Fe~I/H] versus the
EW of each line\footnote{We adopt the observed EW rather than the theoretical
one \citep[log$gf-\theta~\chi_{ex}$,][]{magain84}, following the discussion by
\citet{mucciarelli10}.}; {\em (iii)} the difference between the average [Fe~I/H]
and [Fe~II/H]; and finally by checking that [Fe~I/H] did not change
significantly with wavelength. The adopted atmospheric parameters are listed in
Table~\ref{tab_params} along with the photometric and H$_{\alpha}$ temperatures,
for comparison. We found that, on average, spectroscopic T$_{eff}$ estimates
based on Fe were lower than H$_{\alpha}$ ones by 17$\pm$52~K, and also lower
than photometric ones by 83$\pm$26~K.

\begin{table}
\caption{Atmospheric model parameters.}
\label{tab_params}
\begin{center}
\begin{tabular}{l l c c c c c c} 
\hline\hline
\noalign{\smallskip}
ID$_{WFI}$ & T$_{eff}^{(phot)}$ & T$_{eff}^{(H_{\alpha})}$ & T$_{eff}^{(Fe)}$ &
log~$g^{(Fe)}$ & v$_t^{(Fe)}$ & [M/H] \\
& (K) & (K) & (K) & (dex) & (km s$^{-1}$) & (dex) \\
\noalign{\smallskip}
\hline
\noalign{\smallskip}
503358 & 5600 & 5550 & 5500 & 3.6 & 1.2 & --1.5 \\
503951 & 5600 & 5500 & 5500 & 3.6 & 1.2 & --1.5 \\
507109 & 5600 & 5600 & 5500 & 3.4 & 1.3 & --2.0 \\
507633 & 5700 & 5650 & 5650 & 3.4 & 1.1 & --2.0 \\
512115 & 5600 & 5500 & 5500 & 3.7 & 1.2 & --1.0 \\ 
512938 & 5600 & 5500 & 5550 & 3.2 & 1.4 & --2.0 \\
\hline\hline	       
\end{tabular}	       
\end{center}	       
\end{table}

\subsection{Abundance calculations}
\label{sec_abo}

For most chemical species we computed abundances with the help of the updated
version of the original code by \citet{spite67}. Our reference solar abundance
was that of \citet{gre96}. Once the best set of atmospheric parameters was
chosen for each star (see Section~\ref{sec-param}), we used the new
MARCS\footnote{http://marcs.astro.uu.se/} model atmospheres with standard
composition\footnote{This means [$\alpha$/Fe]=+0.4 for metal-poor stars of
[Fe/H]$<$--1.0 and reaching [$\alpha$/Fe]=0 at [Fe/H]=0, following schematically
the typical halo-disk behaviour of the Milky Way field population.}. We
chose the closest available global model metallicity (taking into account
$\alpha$-enhancement) of the $\omega$~Cen sub-populations, as reported in
Table~\ref{tab_params}.

For all species we computed a 3\,$\sigma$-clipped average of abundances
resulting from each line. For elements that had both neutral and ionized lines,
we computed the weighted (on the number of lines) average of the two ionization
stages to obtain [El/Fe]. We typically rejected lines that had EW$>$100~m\AA,
where the Gaussian approximation could fail, or EW$<$15~m\AA, since the relative
error was too high. For elements with few lines, and where we had to rely mostly
on strong lines, we either performed spectral synthesis, or checked that the
DAOSPEC measurements were not too underestimated by visually inspecting the
spectrum and overlaying the DAOSPEC Gaussian fit on each strong line.

\begin{figure}
\centering
\includegraphics[width=\columnwidth]{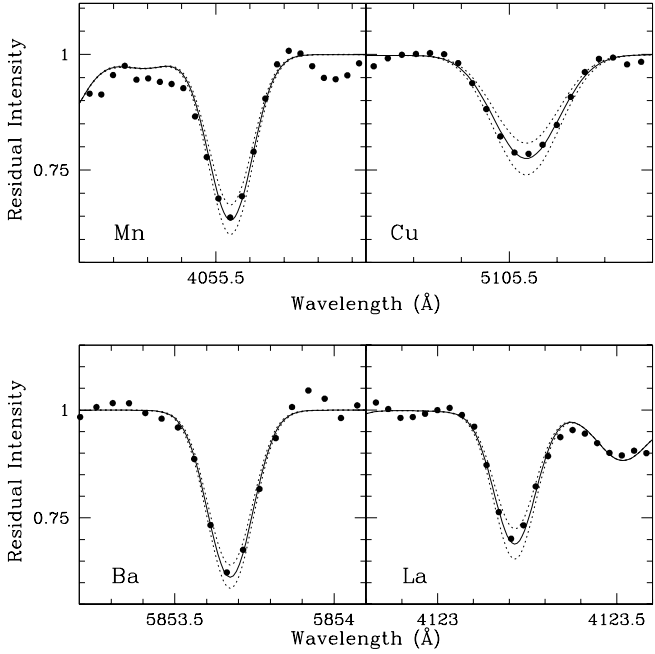}
\caption{Examples of spectral synthesis applied to single lines for star
WFI~512115. Big dots are the observed spectrum, solid lines represent the best
fitting synthetic spectra, computed with all parameters fixed to the values of
Tables~\ref{tab_params} and \ref{tab_abouves}, except for the abundance of the
element under consideration. Spectra computed with the line abundances altered
by $\pm$0.1~dex -- with respect to the best fitting spectrum -- are shown as
dotted lines.}
\label{fig_sint}
\end{figure}

\begin{figure}
\centering
\includegraphics[width=\columnwidth]{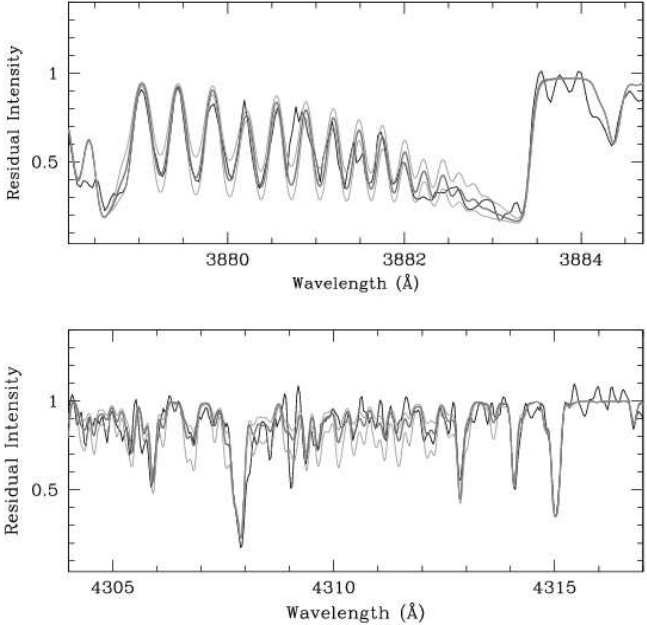}
\caption{Examples of spectral synthesis applied to the CN (top panel) and CH
(bottom panel) molecular bands. Solid black curves represent the observed spectra
for star WFI~512115, thick grey curves represent the best fitting synthetic
spectra, while thin grey lines represent synthetic spectra differing from the
best fit by $\pm$0.2~dex in the N (top panel) and C (bottom panel) abundances.}
\label{fig_cnsynth}
\end{figure}

A few element abundances (Mn, Cu, Ba, and La) were derived with the help of
spectral synthesis, taking into account HFS of single lines when needed (see
also Section~\ref{sec_ew}). We used the
MOOG\footnote{http://verdi.as.utexas.edu/moog.html} \citep{moog} package in
combination with the ``best MARCS models'' above to find the best fitting
spectrum\footnote{To quantify the uncertainty related to the use of two
different codes, one for the EW analysis and the other for spectral synthesis,
we re-analysed star WFI~507633 with MOOG, using the same linelist, EW, atomic
data, atmospheric parameters and MARCS models. We found [Fe/H]=--2.02$\pm$0.02
which is well compatible with the abundance obtained with the Spite code (Table
\ref{tab_abouves}).}. All atmopheric parameters and the abundances of other
blended elements (when present) were kept fixed and only the abundance of the
element of interest was changed until the residuals of the fit were minimized. C
and N abundances were derived by spectral synthesis of the molecular CN and CH
bands at 3880~\AA\  and 4300~\AA, respectively. Examples of line fits with
spectral synthesis are shown in Figures~\ref{fig_sint} and \ref{fig_cnsynth}. 

More in detail, carbon and nitrogen abundances were measured by minimizing the
residuals between grids of synthetic spectra and the observed ones around the CN
and CH bands. In particular, C abundances were derived by fitting the CH G-band
between 4300 and 4340~\AA, including the band heads of the (0-0), (1-1) and (2-2)
bands of the A$^2$$\Delta-$X$^2\Pi$  CH transitions; N abundances were derived
with the CN  B$^2\Sigma^+-$X$^2\Sigma^+$ (0,0) UV band at 3870-3890~\AA. In the
measurement process, we kept the atmospheric parameters and all the atomic
abundances fixed to the values of Tables~\ref{tab_params} and \ref{tab_abouves},
and we used the molecular data from the Kurucz database although, as discussed in
Section~\ref{sec_ew}, a correction of --0.3~dex to the log$gf$ values of C was
necessary. For each star, we computed synthetic spectra around the G-band by
changing the C abundance only, and once the best fit was found, we employed the
found C abundance in the fit of the CN band, where we changed only the N
abundance. The procedure was then iterated a few times, repeating the fit for the
CH band with the newly found N abundance, until the variations in both C and N
were well below 0.1~dex. 

The abundance calculation results are reported in Table~\ref{tab_abouves}, and
are discussed in Section~\ref{sec-res}.

\begin{table*}
\caption{Abundance ratios with random uncertainties (see text).}
\label{tab_abouves}
\begin{center}
\begin{tabular}{l c c c c c c} 
\hline\hline
\noalign{\smallskip}
Element     & WFI~503358      & WFI~503951      & WFI~507109      & WFI~507633      & WFI~512115      & WFI~512938 \\
& (dex) & (dex) & (dex) & (dex) & (dex) & (dex)\\
\noalign{\smallskip}
\hline
\noalign{\smallskip}
[FeI/H]      & --1.70$\pm$0.01 & --1.58$\pm$0.01 & --1.99$\pm$0.02 & --2.05$\pm$0.02 & --1.18$\pm$0.01 & --2.03$\pm$0.02 \\  
$[$FeII/H]   & --1.70$\pm$0.12 & --1.56$\pm$0.09 & --1.95$\pm$0.07 & --1.94$\pm$0.05 & --1.34$\pm$0.05 & --1.99$\pm$0.10 \\ 
$[$Fe/H]     & --1.70$\pm$0.01 & --1.58$\pm$0.01 & --1.98$\pm$0.02 & --2.04$\pm$0.02 & --1.19$\pm$0.01 & --2.03$\pm$0.02 \\ 
$[$Al/Fe]    & +0.13$\pm$(0.05)&  +0.40$\pm$0.29 & --0.12$\pm$(0.05)& +0.14$\pm$0.50 &  +0.80$\pm$0.05 &--0.19$\pm$(0.05)\\
$[$Ba/Fe]    &  +0.57$\pm$0.12 &  +1.10$\pm$0.24 & --0.06$\pm$0.24 & --0.20$\pm$0.16 &  +0.64$\pm$0.12 & --0.33$\pm$0.23 \\
$[$C/Fe]     & --0.02$\pm$0.15 & --0.64$\pm$0.15 &  +0.07$\pm$0.15 & --0.17$\pm$0.15 & --0.94$\pm$0.15 & --0.19$\pm$0.15 \\ 
$[$Ca/Fe]    &  +0.40$\pm$0.03 &  +0.44$\pm$0.03 &  +0.32$\pm$0.05 &  +0.33$\pm$0.04 &  +0.43$\pm$0.08 &  +0.34$\pm$0.03 \\ 
$[$Co/Fe]    & --0.03$\pm$0.19 &  +0.21$\pm$0.12 & --0.14$\pm$0.24 &  +0.08$\pm$0.06 &  +0.37$\pm$0.17 & --0.06$\pm$0.12 \\	  
$[$Cr/Fe]    &  +0.08$\pm$0.23 & --0.08$\pm$0.06 & --0.04$\pm$0.21 & --0.15$\pm$0.10 &  +0.08$\pm$0.04 &  +0.03$\pm$0.17 \\   
$[$Cu/Fe]    &--0.84$\pm$(0.05)&--0.60$\pm$(0.05)&$<$--0.93$\pm$(0.05)&$<$--0.47$\pm$(0.05)&--0.70$\pm$(0.05)&$<$--0.48$\pm$(0.05)\\   
$[$La/Fe]    &  +0.24$\pm$0.06 &  +0.59$\pm$0.05 &  +0.12$\pm$0.08 & --0.13$\pm$0.10 &  +0.18$\pm$0.09 & --0.06$\pm$0.08 \\	
$[$Mg/Fe]    &  +0.46$\pm$0.11 &  +0.41$\pm$0.02 &  +0.35$\pm$0.06 &  +0.45$\pm$0.08 & +0.41$\pm$(0.05)&  +0.55$\pm$0.10 \\   
$[$Mn/Fe]    & --0.78$\pm$0.13 & --0.81$\pm$0.08 & --0.60$\pm$0.11 & --0.41$\pm$0.03 & --0.61$\pm$0.03 & --0.54$\pm$0.11 \\	   
$[$N/Fe]     & --0.12$\pm$0.30 &  +1.36$\pm$0.15 &  +0.57$\pm$0.20 &  +1.23$\pm$0.15 &  +1.26$\pm$0.25 & --0.29$\pm$0.30 \\   
$[$Na/Fe]    &  +0.07$\pm$0.11 &  +0.60$\pm$0.12 &  +0.30$\pm$0.03 & --0.15$\pm$0.33 &  +0.67$\pm$0.19 & --0.42$\pm$0.37 \\   
$[$Ni/Fe]    & --0.10$\pm$0.03 & --0.03$\pm$0.04 &  +0.06$\pm$0.06 &  +0.03$\pm$0.09 & --0.02$\pm$0.04 &  +0.13$\pm$0.08 \\
$[$Sc/Fe]    & --0.03$\pm$0.07 &  +0.20$\pm$0.05 &  +0.10$\pm$0.12 & --0.07$\pm$0.09 & --0.04$\pm$0.12 &  +0.18$\pm$0.14 \\
$[$Si/Fe]    &  +0.38$\pm$0.14 &  +0.46$\pm$0.30 & +0.27$\pm$(0.05)& +0.19$\pm$(0.05)&  +0.35$\pm$0.39 & +0.61$\pm$(0.05)\\   
$[$Sr/Fe]    &  +0.38$\pm$0.04 &  +0.49$\pm$0.05 & --0.13$\pm$0.11 & --0.12$\pm$0.09 &  +0.25$\pm$0.12 & --0.10$\pm$0.09 \\   
$[$Ti/Fe]    &  +0.25$\pm$0.04 &  +0.38$\pm$0.03 &  +0.30$\pm$0.04 &  +0.33$\pm$0.06 &  +0.38$\pm$0.04 &  +0.24$\pm$0.03 \\ 
$[$Y/Fe]     &  +0.58$\pm$0.08 &  +0.71$\pm$0.06 &  +0.31$\pm$0.05 &  ...        &  +0.69$\pm$0.13 & +0.28$\pm$0.72\\ 
\hline\hline
\end{tabular}	        	         			   
\end{center}	        	        				  
\end{table*}

\subsection{Abundance uncertainties}
\label{sec-err}

We estimated the internal, random uncertainty caused by imperfections in the EW
measurement (or spectral synthesis process) and in the line atomic data, as
$\sigma/\sqrt n$, for those elements that had at least two surviving lines after
a 3\,$\sigma$-clipping pass. For elements that relied on one line only,
including those analysed with spectral synthesis, we derived a typical
uncertainty of 0.05~dex by means of the \citet{cayrel88} approximated formula,
and we indicated this typical uncertainty between  parenthesis in
Table~\ref{tab_abouves}.

The uncertainty owing to the continuum normalization procedure can be estimated
from the average spread of the residual spectrum of each star after lines
removal, which is automatically computed by DAOSPEC. For our spectra, this was
$\sim$1\%. According to Figure~2 by \citet{daospec}, this propagates to an
approximate uncertainty in the EW estimates of $\pm$3~m\AA, and finally in an
uncertainty of the order of 0.05~dex in the derived abundances.

Another factor that has a big impact on the abundance ratios is the choice of
atmospheric parameters (Section~\ref{sec-param}). As discussed by
\citet{cayrel04}, T$_{eff}$, log~$g$ and v$_t$ are not strictly independent
parameters when determined with the method of Section~\ref{sec_abo}. Therefore,
the best way to estimate the impact of the parameters' choice on abundance
ratios is to change the most influential parameter, T$_{eff}$, and to
re-optimize the other parameters that naturally re-adjust to accommodate the
temperature change. The difference between abundances calculated with the ``best
model" and with the altered one is a robust estimate of the systematic
uncertainty owing to the choice of stellar parameters. 

We therefore chose our warmest and coolest stars and recomputed their abundances
with models having T$_{eff}$ altered by $\pm$100~K, re-optimizing the other
parameters according to the method described in Section~\ref{sec_abo}. The final
uncertainties are obtained by averaging the absolute abundance differences of the
+100 and --100 altered models, and are reported in Table~\ref{tab_errors}.

The global uncertainty (shown in all Figures from \ref{fig_iron} to
\ref{fig_anti2}) is computed as the sum in quadrature of the random errors, the
uncertainty owing to the continuum placement and the one owing to the choice of
atmospheric parameters. 

\begin{table}
\caption{Uncertanties owing to the choice of stellar parameters.}
\label{tab_errors}
\begin{center}
\begin{tabular}{l c c c} 
\hline\hline
\noalign{\smallskip}
Element & WFI~512115  & WFI~507633  & UVES\\
& (T=5350) & (T=5650) & Average \\
\noalign{\smallskip}
\hline					  
\noalign{\smallskip}			  
[FeI/H]    & $\pm$0.07 & $\pm$0.05 & $\pm$0.06  \\ 
$[$FeII/H] & $\pm$0.05 & $\pm$0.08 & $\pm$0.07  \\ 
$[$Al/Fe]  & $\pm$0.02 & $\pm$0.02 & $\pm$0.02  \\ 
$[$Ba/Fe]  & $\pm$0.04 & $\pm$0.06 & $\pm$0.05  \\ 
$[$Ca/Fe]  & $\pm$0.03 & $\pm$0.04 & $\pm$0.04  \\ 
$[$C/Fe]   & $\pm$0.15 & $\pm$0.15 & $\pm$0.15  \\ 
$[$Co/Fe]  & $\pm$0.03 & $\pm$0.02 & $\pm$0.03  \\ 
$[$Cr/Fe]  & $\pm$0.17 & $\pm$0.04 & $\pm$0.11  \\ 
$[$La/Fe]  & $\pm$0.05 & $\pm$0.05 & $\pm$0.05  \\ 
$[$Mg/Fe]  & $\pm$0.05 & $\pm$0.02 & $\pm$0.04  \\ 
$[$Mn/Fe]  & $\pm$0.03 & $\pm$0.02 & $\pm$0.03  \\ 
$[$N/Fe]   & $\pm$0.20 & $\pm$0.20 & $\pm$0.20  \\ 
$[$Na/Fe]  & $\pm$0.01 & $\pm$0.02 & $\pm$0.02  \\ 
$[$Ni/Fe]  & $\pm$0.02 & $\pm$0.01 & $\pm$0.02  \\ 
$[$Sc/Fe]  & $\pm$0.06 & $\pm$0.06 & $\pm$0.06  \\ 
$[$Si/Fe]  & $\pm$0.02 & $\pm$0.02 & $\pm$0.02  \\ 
$[$Sr/Fe]  & $\pm$0.09 & $\pm$0.15 & $\pm$0.12  \\ 
$[$Ti/Fe]  & $\pm$0.02 & $\pm$0.05 & $\pm$0.04  \\ 
$[$Y/Fe]   & $\pm$0.05 & $\pm$0.06 & $\pm$0.06  \\ 
\hline\hline	     		         			     			     			        
\end{tabular}	        	         			   
\end{center}	        	        				  
\end{table}

\subsection{A comparison with NGC~6397}
\label{sec_6397}

Usually, to test the goodness of an abundance analysis, a comparison with the
Sun or Arcturus is performed. In our case, because our linelist was optimized
for metal-poor subgiants, we would be performing an abundance analysis of the
Sun relying only on strong lines, with EW$>$100~m\AA. These strong lines are
rejected when analysing our UVES metal-poor stars, because at this resolution
they start to significantly deviate from the Gaussian shape \citep[see Figure~7
by][]{daospec}. 

\begin{table}
\caption{Comparative analysis of NGC 6397.}
\label{tab_6397}
\begin{center}
\begin{tabular}{l c c c c l} 
\hline\hline
\noalign{\smallskip}
Star & T$_{eff}$  & log$g$  & v$_t$ & $\Delta$[Fe/H]$^{a}$ & Notes\\
& (K) & (dex) & (km/s) & (dex) & \\
\noalign{\smallskip}
\hline					  
\noalign{\smallskip}			  
699    & 5500 & 3.4 & 1.3 &  +0.00 & \citet{gratton01} \\
       & 5400 & 3.3 & 1.8 & --0.10 & \citet{korn07} \\
793    & 5500 & 3.4 & 1.3 & --0.05 & \citet{gratton01} \\
       & 5500 & 3.4 & 1.6 &  +0.06 & \citet{korn07}  \\ 
206810 & 5500 & 3.4 & 1.3 &  +0.09 & \citet{gratton01} \\
       & 5500 & 3.4 & 1.8 &  +0.07 & \citet{korn07}  \\              
\hline\hline
\multicolumn{6}{l}{$^a$Differences in the sense: ours minus literature, with
same parameters.}\\	     		         			     			     			        
\end{tabular}	        	         			   
\end{center}	        	        				  
\end{table}

Therefore, while the general method was already tested on the Sun by
\citet{p10}, we preferred to compare our present results to another well studied
cluster, such as NGC~6397. We downloaded UVES archive spectra of three
sub-giants (stars 206810, 669, and 793) in NGC~6397, published by
\citet{gratton01} and later re-analysed by \citet{korn07}, which have similar
temperatures, gravities, and metallicities as the three most metal-poor SGB
stars in our sample. We re-analysed them adopting the same parameters as
\citet{gratton01} and as \citet{korn07}, respectively, as detailed in
Table~\ref{tab_6397}. Obviously, the differences in the resulting [Fe/H] are
negligible within the quoted uncertainties. We note in passing that the major
difference between the \citet{gratton01} and \citet{korn07} abundance
determinations for these three stars lies in the v$_t$ determination, which is
approximately 0.5~km s$^{-1}$ higher in the \citet{korn07} paper. This
difference alone is probably enough to justify the different abundances found in
the two studies.

\begin{figure}
\centering
\includegraphics[angle=270,width=\columnwidth]{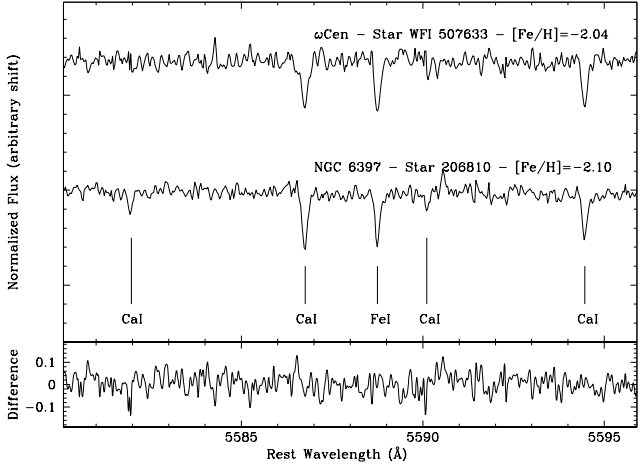}
\caption{Comparison between the spectra of our UVES star WFI~507633 (topmost
spectrum in the top panel) and star 206810 in NGC~6397 from the UVES archive
(bottom spectrum in the top panel). The bottom panel shows the difference between
the two spectra.}
\label{fig_6397}
\end{figure}

Figure~\ref{fig_6397} compares the spectra of WFI~507633 and star 206810 in
NGC~6397. The parameters of these stars are very similar, and indeed the calcium
and iron lines are virtually identical, as confirmed by subtracting one spectrum
from the other, where only noise remains. We are then confident that our
analysis is correct and consistent with the one by \citet{gratton01}, within the
estimated uncertainties.

\section{Abundance results}
\label{sec-res}

In the following sections we present and discuss abundance ratios of the
measured elements. To compare with homogeneous literature sets, we considered
preferentially large ($\geq$10 stars) samples of high-resolution measurements of
stars belonging to $\omega$~Cen. Unfortunately, there are no high-resolution
studies of subgiants\footnote{Although \citet{villanova07} study stars in the
same evolutionary phase as our targets, their resolution is lower
(R$\simeq$6000).}, so we resorted to red giants surveys
\citep{norris95,smith00,johnson09}. This will help us to cross-identify
sub-populations in the SGB region with those already identified on the RGB. For
some species, there were no large literature surveys available, so we included
articles based on smaller samples ($<$10 stars). The largest sample available to
date is the one by \citet{johnson10}, which contains more than 800 stars, but at
a moderate resolution (R$\simeq$18\,000). Their abundance ratio plots are more
populous, but also obviously more scattered compared to higher-resolution
studies, so we decided not to plot them in our Figures~\ref{fig_iron} to
\ref{fig_anti2}, except for their stars at [Fe/H]$<$--1.9~dex, which are not as
well sampled in other less populous studies. We will also mention the abundace
ratios by \citet{villanova10}, a study of 38 stars observed in the external
regions of $\omega$ Cen. Given the diversity of comparison studies, even if we
reported all literature values to our solar abundance and log$gf$ system (using
lines in common) when possible, residual zeropoint differences among different
studies will surely still be present.

\subsection{Iron}
\label{sec-iron}

\begin{figure} 
\centering \includegraphics[width=\columnwidth]{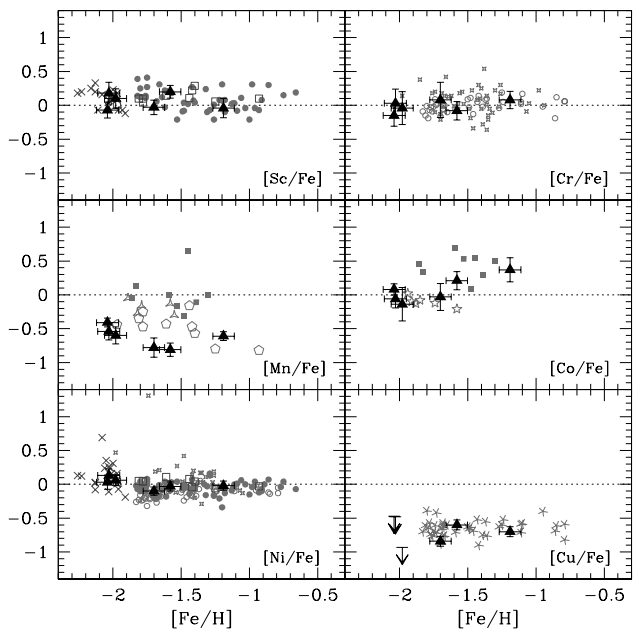}
\caption{Abundance ratios of iron-peak elements. Our measurements are marked
with filled black triangles: error bars include random errors, continuum
placement errors and errors owing to the choice of atmospheric parameters (see
text). Grey symbols mark literature measurements for red giants: filled circles
are from \citet{johnson09}, empty pentagons from \citet{cunha10}, empty circles
from \citet{norris95}, empty triangles from \citet{cohen81}, filled squares from
\citet{gratton82}, empty stars from \citet{francois88}, asterisks from
\citet{cunha02}, crosses from \citet{johnson10}, four-pointed stars from
\citet{villanova10}.} 
\label{fig_iron} 
\end{figure}

Our abundance ratios for iron and iron-peak elements are shown in
Figure~\ref{fig_iron}. We note that three of our six stars lie around
[Fe/H]$\simeq$--2.0~dex, and that they do not correspond to any of the major
sub-populations identified on the RGB, which are more metal-poor than any red
giant studied with high-resolution spectroscopy before. The only exceptions are
star ROA~213 by \citet{smith00}, which has [Fe/H]=--1.97 dex, star 85007 by
\citet{villanova10} at [Fe/H]=--1.98 dex, and a small group of 25 stars (out of
more than 800) below [Fe/H]$\simeq$--2.0~dex in Figure~10 by \citet{johnson10}.
In their figure, the bulk populations have higher metallicities, with an abrupt
drop in numbers below [Fe/H]$\simeq$--1.9~dex, approximately. We tentatively
classify our three stars as a separate subgroup -- containing few stars -- that
we name VMP (from very metal-poor, see also Section~\ref{sec_first}), and we
will consider ROA~213 by \citet{smith00} and the 25 stars by \citet{johnson10}
as the VMP counterparts on the RGB in the following discussion. The other three
targets apparently fall into the metal-poor and metal-intermediate populations
\citep[MP and MInt, according to the classification by][]{p00,sollima05a}. We
will discuss population identifications in more detail in Section~\ref{sec_who}.

We note that all comparisons in Figure~\ref{fig_iron} are between stars in
different evolutionary phases (our targets are subgiants, while those in the
literature are always giants). Therefore, the uncertainty on the zeropoint --
especially as far as [Fe/H] is concerned -- cannot be well quantified (see also
Section~\ref{sec-err}). Possible factors that cause systematically discrepant
abundances between giants and subgiants are 

\begin{itemize}
\item{there is still no consensus for the NLTE corrections to Fe\,I abundances,
ranging from negligible \citep{gratton99}, to approximately 0.03--0.05~dex
\citep{korn07}, to +0.3~dex \citep{thevenin99} for stars of the type presented
here;} 
\item{the effect of diffusion could already be present for these subgiants,
lowering abundance by an amount that could be negligible \citep{gratton01} or
of the order of 0.1--0.2~dex \citep{korn07}; however, diffusion is heavily
influenced in all models by the amount of turbulent mixing, a phenomenon that
is poorly constrained at present;} 
\item{there are also possible granulation inhomogeneities, which apparently go
in the opposite direction compared to the above effects \citep{shchukina05};} 
\item{finally, two studies that analysed these phenomena in NGC~6397
\citep{gratton01,korn07} gave very different abundance results (see
Section~\ref{sec_6397}), so even from the observational point of view it is
quite difficult to provide reliable constraints to these phenomena.}
\end{itemize}

In spite of all these problems, we have the advantage that we do not use only
[Fe/H], but we can also compare [El/Fe] abundance ratios -- which should be more
robust to measurement uncertainties and to variations caused by these physical
processes -- with a vast literature, to effectively cross-identify SGB and RGB
populations.

\subsection{Iron-peak elements}
\label{sec-fepeak}

While \citet{norris95} provided Cr and Ni measurements for 40 red giants,
\citet{johnson09} gave Ni measurements for 66 red giants and Cr for a smaller
subset. Cr and Ni measurements are also provided by \citet{villanova10} for
30-35 stars. We took the \citet{cunha02} copper measurements for the 40 red
giants already analysed by \citet{norris95}\footnote{The Cu measurements of six
stars by our group \citep{p02} are in substantial agreement with the ones by
\citet{cunha02}, so we do not plot them in the various Figures.}. For stars with
[Fe/H]$<$--1.9~dex we used the \citet{johnson10} Sc and Ni measurements. No
literature measurements in large samples of red giants were found for Co and Mn,
except for the recent Mn study by \citet{cunha10}, so we also compared with five
giants by \citet{cohen81}, eight giants by \citet{gratton82}, and six giants by
\citet{francois88}. 

There is general agreement for all iron-peak elements between our measurements
for SGB stars and the literature RGB ones. Two odd elements, Sc and Co, would
give better results if hyperfine splitting (HFS) was taken into account, but
since they do not give any additional information with respect to Fe,
traditionally no auhor has employed spectral synthesis to measure them. As a
result, all ratios of Sc and Co in Figure~\ref{fig_iron} -- including ours --
are on average above solar. Our Sc values fall exactly on top of literature
measurements, and our Co values for the three most metal-rich SGB stars lie
between the \citet{gratton82} and the \citet{francois88} measurements. It is
known that large NLTE corrections are required for Co \citep{bergemann10}, which
explains the rising trend observed in all cited studies. A slight underabundance
can be noticed in the chromium ratios for the three most metal-poor stars, but
NLTE effects should again be the cause for it \citep{bergemann10}. It is
interesting to note that [Ni/Fe] tends to be slightly higher for the VMP stars
than for the remaining three UVES targets by 0.1--0.2~dex, and that a similar
trend can be noticed in Figure~10 by \citet{johnson10}, who do not comment on
this effect.

Mn and Cu are two very interesting elements with a dedicated vast literature
because, although they belong to the iron-peak, they behave differently from
other iron-peak elements \citep[see][for a classical review]{mcwilliam97}. In
particular, Cu in $\omega$~Cen has been first measured by \citet{p02} on six RGB
stars, and later studied in detail by \citet{cunha02}, who found it
underabundant similarly to field stars. Finally, it was theoretically modelled
by \citet{romano07}, who found massive stars as the most likely producers of Cu,
in agreement with previous studies \citep{bisterzo04}. While we find only upper
limits for the three most metal-poor stars (Figure~\ref{fig_iron}), our
measurements agree with those by \citet{cunha02}. 

\begin{figure}
\centering
\includegraphics[bb=20 150 590 560,clip,width=\columnwidth]{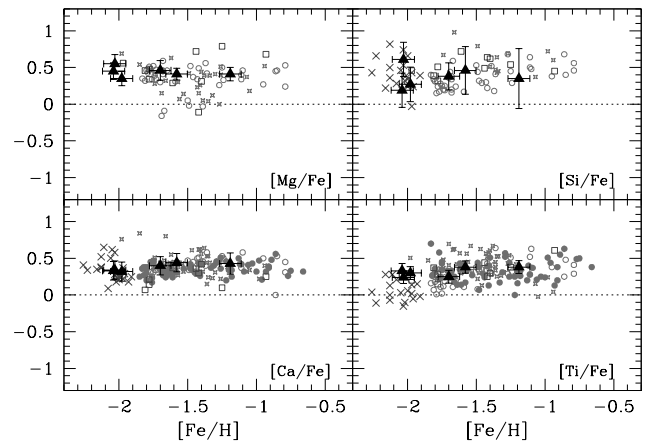}
\caption{Abundance ratios of $\alpha$-elements. Symbols are the same as in
Figure~\ref{fig_iron}.}
\label{fig_alfa}
\end{figure}

Manganese instead was found to be extremely underabundant in our six stars. The
determinations by \citet{cohen81} and \citet{gratton82} appear only slightly
subsolar, but none of these estimates takes into account HFS. A better agreement
was found with the measurements by \citet{cunha10}, based on the 10 stars by
\citet{smith00}, and derived with a complete HFS analysis, of which we plot the
LTE resulting abundances in Figure~\ref{fig_iron}. Our determinations, as those
of \citet{cunha10}, are lower than the values found for field stars
\citep{mcwilliam97} or metal-poor stars \citep{cayrel04} around
[Fe/H]=--2.0~dex. In particular, three of the five analysed lines belong to the
4000~\AA\  resonance triplet which, according to \citet{cayrel04} gives
systematically lower abundances by approximately 0.4~dex. Even rejecting these
three lines, we found a [Mn/Fe]$\simeq$--0.7~dex for each of our six stars
(Figure~\ref{fig_iron}). The (marginal) discrepancy between our results and
those by \citet{cunha10} could arise because we used the Mn lines around
4000~\AA, while they use the lines around 6000~\AA. The general result -- that
$\omega$~Cen has lower Mn abundance than field and GGC stars -- suggests that
the source of Mn production in $\omega$Cen should be low-metallicity supernovae,
either of type II or Ia \citep{cunha10}.

\subsection{$\alpha$-Elements}
\label{sec-alfa}

It was not possible to measure the oxygen lines, or to put meaningful upper
limits to the oxygen abundance in these subgiants. We were able to measure Mg,
Si, Ca and Ti (Figure~\ref{fig_alfa}): all four elements were present in the
\citet{norris95} and \citet{smith00} studies, but we plotted also the Ca and Ti
more recent measurements by \citet{johnson09} and the Si, Ca, and Ti
measurements for stars with [Fe/H]$<$--1.9~dex by \citet{johnson10}. The
measurements by \citet{villanova10} also include all four $\alpha$-elements
studies here.

All measurements are slightly dispersed for Si because only a handful of lines
is generally available, while for the other three elements they show a
reasonably small spread. The three VMP stars do not seem to show any difference
from the other three SGB stars in their $\alpha$-enhancement, with a weighted
average of [$\alpha$/Fe]=0.36$\pm$0.08~dex for all elements in all six stars.
Literature measurements are quite scattered for Mg, but for RGB stars the Mg
lines are very strong and difficult to measure, and Mg also anti-correrelates
with Na, for example, so some spread should be expected. Apart from this
exception, there is very good agreement among various literature determinations
for giants, and with our measurements for subgiants. In particular, we note that
RGB stars with [Fe/H]$<$--1.9~dex are always very close to our three VMP
subgiants. The NLTE effects for all our SGB stars should be around
$\simeq$0.1~dex in size \citep{gratton99}, for Mg, but not all the lines
analysed here were also analysed by \citet{gratton99}, so we prefer not to apply
any correction.

\begin{figure}
\centering
\includegraphics[bb=20 150 590 560,clip,width=\columnwidth]{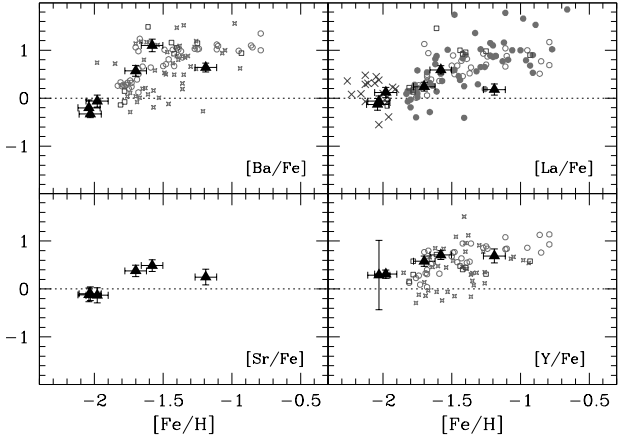}
\caption{Abundance ratios of heavy elements. Symbols are the same as in
Figure~\ref{fig_iron}.}
\label{fig_esse}
\end{figure}

The $\alpha$-enhancement of VMP, metal-poor, and intermediate stars in
$\omega$~Cen is typical of the field stars of the Galactic halo, which is
indicative of a chemical enrichment dominated by SNe~II. 

\subsection{Heavy elements}
\label{sec-esse}

We were able to measure Ba, La, Sr, and Y. For Ba and La we used spectral
synthesis (see Sections~\ref{sec_ew} and \ref{sec_abo}) to take into account
HFS. We plot the abundance ratios in Figure~\ref{fig_esse}, along with the
measurements by \citet{norris95}, \citet{johnson09}, \citet{villanova10}, and
\citet{johnson10} for stars with [Fe/H]$<$--1.9~dex. The log\,$gf$ used by
\citet{norris95} are now outdated, which is why, using the stars in common
between literature studies \citep[including][]{vanture94,smith00,p03}, we
estimated that all their ratios needed to be raised by $\simeq$0.4--0.5~dex --
depending on the element -- to compare with more recent studies. The
\citet{norris95} data appearing in Figure \ref{fig_esse} are already corrected
for this effect. Also, the solar Ba abundance by \citet{villanova10} was 2.31,
while we use 2.13, and their data are corrected for this difference in
Figure~\ref{fig_esse} as well.

The heavy s-process elements Ba and La generally agree with literature
estimates. In particular, the VMP stars appear to have a lower average s-process
enhancement, compatible with zero, as confirmed by the La measurements by
\citet{johnson10}. Thus, no s-process enrichment by AGB stars seems to have
polluted the VMP star in the RGB, or in the SGB. This is supported by our Sr and
Y measurements, which mainly agree with the literature measurements and which
also have a low enhancement, compatible with zero for our VMP stars. 

It is also interesting that the most metal-rich star in our sample, WFI~512115,
appears to have a slightly lower s-process enhancement than other stars measured
in the literature at similar metallicity. If confirmed by larger samples of
metal-rich stars, this could point out that the s-process enrichment by AGB
stars in $\omega$~Centauri was not completely homogeneous, as supported also by
the small group of stars in the \citet{norris95} sample, around
[Fe/H]$\simeq$-1.7, which appears to have [Ba/Fe] 0.2--0.3~dex higher than other
stars at similar metallicity. A larger spread was found by \citet{villanova10}
in s-process elements than in other studies presented here, which they explain
by assuming that a bimodality in these elements might be present at low
metallicity, supporting the mentioned effect in the \citet{norris95} data. Our
three stars always share the same s-process enhancement, but larger samples are
of course needed to see if the supposed bimodality, or higher spread, extends to
-2.0 dex stars as well. Finally, there is a large fraction of stars with solar
Ba and Y enhancement in the \citet{villanova10} sample at all metallicities,
which are not present in any of the other studies of red giants in $\omega$ Cen,
and it is not clear yet if this is because of an intrinsic feature of stars in
the outskirts of the stellar system, or a spurious measurement effect.

\subsection{Anti-correlations}
\label{sec-anti}

\begin{figure}
\centering
\vspace{-0.8cm}
\includegraphics[angle=270,width=\columnwidth]{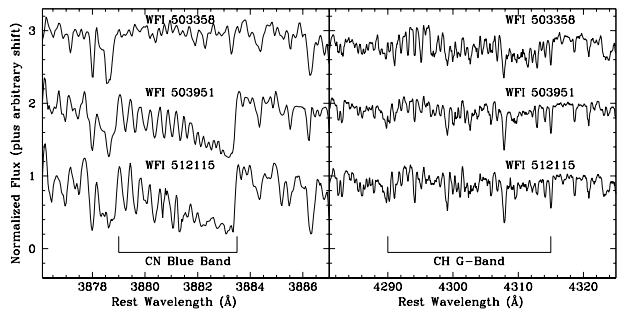}
\vspace{-1cm}
\caption{Smoothed spectra of the three most metal-rich UVES targets around the
CN and the CH blue bands.}
\label{fig_antisp}
\end{figure}

Figure~\ref{fig_antisp} shows the CH and CN band regions around 3880 and
4300~\AA , respectively, for the three most metal-rich UVES stars. Clearly, two
of them (namely WFI~503951 and 512115) show a deep CN band and almost no signal
in the CH band. On the opposite, star WFI~503358 shows almost no CN band, and
the CH band appears slightly deeper than in the other two stars. This evidence
is not so clear in the three most metal-poor UVES stars, since the diatomic CN
molecular band is much shallower. 

\begin{figure}
\centering
\includegraphics[bb=20 150 590 560,clip,width=\columnwidth]{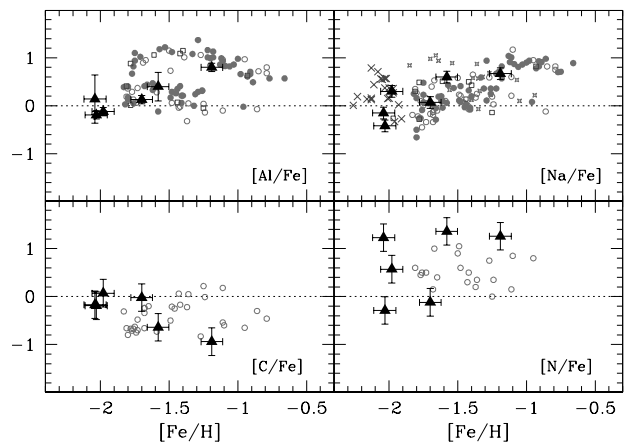}
\caption{Abundance ratios of Al, Na, C, and N for  the six UVES stars. Symbols
are the same as in Figure~\ref{fig_iron}.}
\label{fig_anti1}
\end{figure}

We also have atomic lines of other elements that are found to (anti-)correlate
within Galactic globular clusters, namely Mg, Na, and Al, besides C and N. Their
abundance ratios are plotted in Figures~\ref{fig_alfa} and \ref{fig_anti1}. An
initial observation is, looking at the literature abundance ratio plots
\citep{norris95,johnson09,johnson10,marino10} that [Al/Fe] and [Na/Fe] appear
clearly bimodal at all metallicities (with the possible exception of the most
metal-rich stars around [Fe/H]$\simeq$--0.6~dex), with two distinct sequences
around [Al/Fe]$\simeq$0 and [Al/Fe]$\simeq$1~dex in the case of aluminium, and
the same bimodality is visible, although less clearly, in the C and N ratios. In
our Figure \ref{fig_anti1} the Na data are confused because the
\citet{villanova10} data have an opposite trend to all the other literature data
plotted, with Na decreasing with metallicity instead of increasing with it. We
found no explanation for this trend. Apart from this, our data follow the
literature trends reasonably well, with a similarly large scatter for [N/Fe]. We
did not apply NLTE corrections following a reasoning similar to that of Mg in
Section~\ref{sec-alfa}, but they should be again around 0.1~dex for these
subgiants \citep{gratton01}.

\begin{figure}
\centering
\includegraphics[width=\columnwidth]{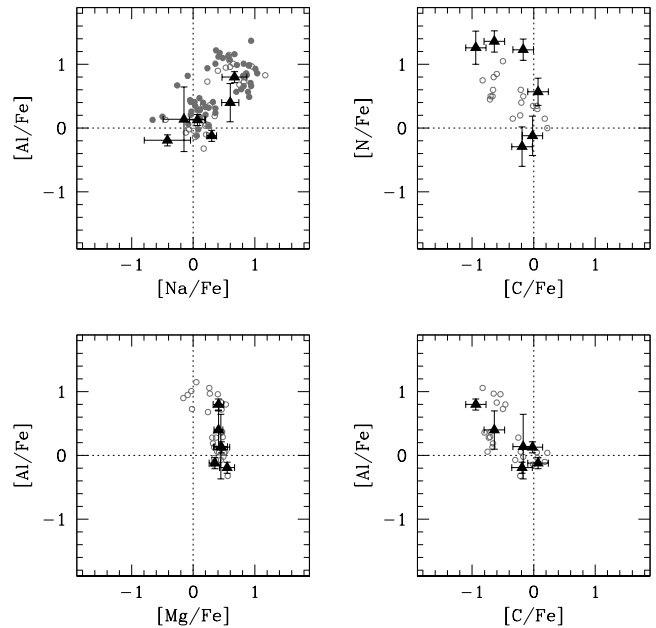}
\caption{Anti-correlation plots for the six UVES stars. Symbols are the same as
in Figure~\ref{fig_iron}.}
\label{fig_anti2}
\end{figure}

Figure~\ref{fig_anti2} reports on the usual (anti-)correlation plots. A clear
and bimodal Na-Al correlation is seen in the literature data, as well as the C-N
anti-correlation. Our data compare well with the literature in both cases. Even
the most difficult Mg-Al anti-correlation is clearly visible in the literature
data, although not as clearly in our own data for the six UVES stars. Finally,
we also report the C-Al anti-correlation in Figure~\ref{fig_anti2}. 

In summary, two of the three most metal-rich stars show clear signs of CNO
pollution, as well as a clear Na-Al anti-correlation, and this agrees with all
past literature data on the subject. Our data for the three VMP stars are
instead not conclusive about the presence of (anti-)correlations in the VMP
population as a whole. We have two stars (WFI~507109 and 512938) with primordial
composition and one star (WFI~507633) that seems to have high [N/Fe], but solar
[C/Fe]. Literature measurements seem to point towards absent or reduced 
(anti-)correlations in VMP stars. The 25 stars by \citet{johnson10} with
[Fe/H]$<$--1.9~dex tend to all have high sodium, and a much lower dispersion
than the other $>$800 stars in their sample. But there are no Al, Mg, C, or N
measurements in their study. On the other hand, \citet{marino10} show that for
their red giants with [Fe/H]$<$--1.8~dex, the Na-O anti-correlation appear less
extended than that for other metallicity groups in $\omega$~Cen, but we do not
know how the situation stands for [Fe/H]$<$--1.9~dex only. 

While a complete discussion on the anti-correlations among light elements in
$\omega$~Centauri is beyond the scope of the present paper -- for the moment we
just want to cross-identify our SGB stars with the RGB populations -- we confirm
that whatever the cause of anti-correlations in GC, it must have been active on
each sub-population of $\omega$~Cen, possibly excluding only the most metal-poor
one, at [Fe/H]$\simeq$--2.0~dex, and the most metal-rich one \citep[where all
stars above --1.0~dex appear polluted,][]{marino10}. The presence or absence of
(anti-)correlations in some sub-populations of $\omega$~Cen is a powerful tool
to understand its chemical evolution \citep[see][and the discussion in
Section~\ref{sec_first}]{p03,carretta10}.

\section{Discussion}
\label{sec-disc}

While comparisons between subgiants and giants are not entirely free from
problems \citep[see e.g.,][and references therein]{bonifacio09}, we measured
abundance ratios of several species, in good agreement with past literature
determination for RGB stars. We found that three of our targets seem to belong
to a separate population with typical [Fe/H]$\simeq$--2.0~dex,
[$\alpha$/Fe]$\simeq$+0.35~dex, and an $s$-process enhancement compatible with
zero. Of the three remaining stars, two are consistent with the abundance ratios
of the RGB-MP \citep[nomenclature by][]{p00}, and the last star  appears to
belong to the RGB-MInt population \citep[nomenclature by][]{p00} or, more
specifically, to the RGB-MInt2 \citep[nomenclature by][]{sollima05a}.

With this identification as our compass, we will try in the following sections
to contribute to the understanding of the SGB of $\omega$~Cen on a few crucial
topics.


\subsection{The first stellar generation in $\omega$~Cen?}
\label{sec_first}

As discussed in Section~\ref{sec-iron}, the three most metal-poor stars at
[Fe/H]$\simeq$--2.0~dex appear to belong to a separate sub-population, which we
termed VMP (for very metal-poor). Stars as metal-poor as these were found in the
past in small amounts, but were not considered as a separate sub-population per
se. In particular, several (unbiassed with respect to metallicity) studies,
aimed at deriving the metallicity distribution of $\omega$~Cen either through
photometry \citep{frinchaboy02,sollima05a,calamida09}, low-resolution
spectroscopy of red giants \citep{norris96,suntzeff96}, or subgiants
\citep{sollima05b,stanford06,villanova07} found stars as poor as --2.0~dex or
lower. Some high-resolution abundance studies also found again a few VMP stars
\citep{smith00,johnson09,johnson10,marino10,villanova10}, and in general there
appears to be an abrupt termination of the main MP population around
[Fe/H]$\simeq$--1.9~dex, with a sparse group of stars around
[Fe/H]$\simeq$--2.0~dex. This effect is clearly visible in Figure~10 by
\citet{johnson10}. This clean behaviour justifies the definition of VMP as a
new, separate sub-population in $\omega$~Cen. 

To estimate the fraction of such a minority population is not easy. From the
cited studies we estimate that it should be at most 5\% of the entire stellar
content of $\omega$~Cen. Some more support to the existence of this small VMP
component comes from the recent work by \citet{bellini10}, who used exquisite
ACS photometry to reveal additional sub-structure in the SGB region of
$\omega$~Cen. In particular, the upper SGB branch \citep[branch A in the
nomenclature of][]{villanova07} appears split into two sub-branches in
\citet{bellini10}. The fact that three out of three of our upper SGB targets
turned out to have [Fe/H]=--2.0~dex can thus be explained, because we chose them
on the upper evelope of the upper branch, which turns out to be separated from
the other branches, and which we can safely consider to be made of
[Fe/H]=--2.0~dex stars, at least in the external region of $\omega$~Cen that we
are sampling here (see Figure~\ref{fig_area}).

Our abundance ratios, based on high-resolution spectroscopy, allow us to
hypothesize that these stars must be the best candidate remnant of the
primordial population in $\omega$~Centauri, enriched primarily by type II SNe
(given its [$\alpha$/Fe]$\simeq$+0.35~dex), and most probably free from severe
pollution by AGB stars. This last statement is supported by the (almost) solar
$s$-process ratios (Figure~\ref{fig_esse}). Also, while three stars are too few
to rule out the presence of anti-correlations in this population, literature
studies \citep{smith00,johnson09,johnson10,marino10} suggest that
(anti-)correlations should be absent or reduced in their extension for VMP
stars. It would be {\em extremely} interesting to study C, N, Mg, Na, and Al in
larger samples of SGB stars in $\omega$~Cen. Indeed, although $\omega$~Cen is
commonly considered as the remnant of a dwarf galaxy accreted a long time ago by
the Milky Way, all the sub-populations identified so far show clear
anti-correlations (Figures~\ref{fig_anti1} and \ref{fig_anti2}) at all
metallicities, except for -- possibly -- the VMP and RGB-a stars
\citep{marino10}. Field populations of dwarf galaxies do not normally show any
anti-correlation, which is only found in globular clusters \citep{gratton04}.
Therefore, the absence (or existence) of anti-correlations in the VMP component
could give us evidence for (or against) the dwarf galaxy hypothesis for the
origin of $\omega$~Centauri, as discussed also by \citet{p03} and
\citep{carretta10}.

Summarizing, given the chemical properties of the VMP stars studies here and in
the cited high-resolution studies, we conclude that they belong to a small and
distinct sub-population, which appears to be the best-candidate (remnant)
population of the first stellar generation in $\omega$~Cen, but could also -- if
the presence of (anti)-correlations will be excluded -- be the field population
remnant of its hypothesized parent galaxy.

\subsection{SGB populations puzzle: who is who} 
\label{sec_who}

\begin{figure*}
\centering
\vspace{-4cm}
\includegraphics[angle=270,width=\textwidth]{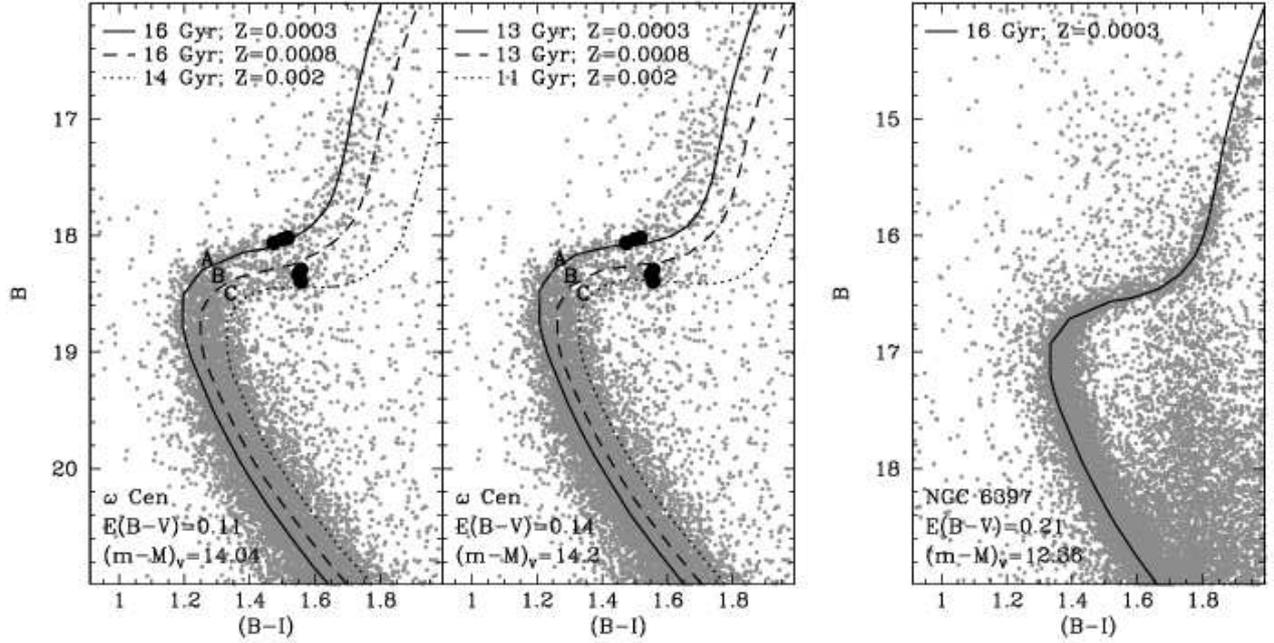}
\vspace{-0.5cm}
\caption{An example of isochrones fit (see text) to determine age differences
among the three sub-populations studied in this paper. Left panel: the
literature values of E(B--V) and (m-M)$_V$ for $\omega$~Cen are used to derive
age differences; the oldest populations turns out to be 16~Gr old. Centre panel:
a cosmological age of 13~Gyr is forced for the oldest populations, altering
E(B--V) and (m-M)$_V$ but with no significant effect on the resulting age
differences; the shape of the isochrones is now in worse agreement with the
data. Right panel:  NGC~6397 is shown for comparison. The letters A, B, and C,
which appear in the two leftmost panels, indicate the approximate location of
the SGB branches as defined by \citet{villanova07}.}
\label{fig_ages}
\end{figure*}

Apart from the notation used above, which separates the $\omega$~Cen
sub-populations by metallicity in VMP, MP, MInt (in turn divided in Int1, Int2,
and Int3), and MR (or RGB-a and SGB-a) proposed by \citet{p00},
\citet{ferraro04} and \citet{sollima05a}, we will also use the nomenclature by
\citet{villanova07} in this Section, who photometrically divide the SGB into
four sub-branches (A, B, C and D) from the upper SGB envelope down to the SGB-a.
This classification is also put into evidence in the two leftmost panels of
Figure~\ref{fig_ages}. 

The whole difficulty in the study of the SGB of $\omega$~Cen is to
cross-correlate the five sub-populations defined photometrically and
spectroscopically along the RGB with the four or five \citep[and more,
see][]{bellini10} sub-branches that are visible on the SGB. This is because,
while on the RGB there are exquisite high-resolution studies that can link
photometric features with chemical abundance patterns, this is not yet possible
for the SGB. One important observation suggested by past attempts to link SGB
and RGB populations is that the nicely combed structure that appears in the RGB
and the similarly nicely combed one that appears on the SGB could be somehow
scrambled and mixed, as pointed out by \citet{villanova07}. While on the RGB the
dominating factor that separates populations is metallicity, on the SGB it must
be age (together with C, N, O, and He), and this could complicate things
significantly if there were no clear (monotonic increasing) age-metallicity
relation in $\omega$~Cen. The uncertainties involved in the past low-resolution
spectroscopic studies such as \citet{sollima05b} and \citet{villanova07} were
apparently not sufficient to give a final answer to the problem. The present
study, on the other hand, while having higher precision in the abundance
determination, can only rely on a limited number of stars. 

\subsubsection{The VMP population} 

Concerning our newly defined VMP population, we must assume that it should lie:
{\em (i)} on the bluest egde of the RGB colour distribution, because of its low
metallicity, and {\em (ii)} on the upper edge of the SGB, and in particular on
the upper edge of branch A by \citet{villanova07}, where we find three stars out
of three at [Fe/H]$\simeq$--2~dex. This is supported by Figure~12 by
\citet{bellini10}, where for the first time the upper SGB appears split into two
separate branches, and where the lower base of the RGB shows a clearly bluer
sub-sequence made of a small number of stars. These features are the most likely
photometric counterparts of the VMP population found here. 

\subsubsection{The MP population} 

The connection between the SGB-MP and the RGB-MP is not clear yet. While {\em
some other population} should be present together with the VMP to account for the
number of stars in the upper branch, or branch A, it is not clear if this is
really the counterpart of the RGB-MP, because the RGB-MP contains at least 30\%
(and likely more) of the stars in $\omega$~Cen. We cannot speculate too much with
the data in hand -- the only certain fact is that the two MP stars in this paper
(WFI~503358 and 503951) both lie lower than branch A, possibly on branch B or
even C -- but we see two possible solutions: 

\begin{itemize} 

\item{One possibility -- supported by our analysis of stars WFI~503358 and
503351 -- is that the RGB-MInt1 stays on branch A (lower half) together with the
VMP, and the RGB-MP stays instead on branch B. This would allow for a better
agreement between the population fractions estimated by \citet{sollima05a} and
\citet{villanova07}, while the uncertainties of both low resolution studies are
probably large enough to accommodate such a switch.}

\item{The other obvious interpretation has been already mentioned by
\citet{villanova07}, i.e., that MP stars lie on both branches A (lower) and B,
mixed with part of the metal-intermediate populations. }

\end{itemize} 

\subsubsection{The MInt populations} 

Our star WFI~512115, which chemically belongs to the MInt population (and, more
specifically, to the MInt2) also appears to lie either on branch B or -- most
probably -- on branch C. Combining this evidence with the one presented by both
\citet{sollima05b} and \citet{villanova07}, we tentatively conclude that the
most likely photometric counterpart of the MInt2 population should be branch C.
The presence of a few RGB-MP stars on branch C still cannot be ruled out, and it
could be either caused by some scatter in the low-resolution abundances by
\citet{villanova07}, or by photometric errors in the WFI reference photometry
\citep{p00,p03}, or even by a different SGB shape for the SGB-MP
\citep{bellini10}. We also speculate that the RGB-MInt3 should correspond to a
sequence lying between branches C and D, as clearly indicated by
\citet{sollima05b} in their Figure~4, and mentioned in passing by
\citet{villanova07}. 

\subsubsection{The MR or SGB-a population}

While the cross-identification between the RGB-a \citep{p00,sollima05a}, the
SGB-a \citet{ferraro04, sollima05b} and branch D \citep{villanova07} appears
unquestionable, there is still some debate about its metallicity, ranging from
[Fe/H]$\simeq$--1.1, derived from low-resolution spectroscopy of SGB stars by
\citet{villanova07} to [Fe/H]$\simeq$--0.6~dex, derived from both high-resolution
spectroscopy of RGB stars \citep{p00} and low-resolution spectroscopy of SGB
stars \citep{sollima05b}. The metallicity of the SGB-a will be the subject of a
following paper based on GIRAFFE data, but it is interesting to note here that
\citet{bellini10} found the SGB-a (or branch D) split into two sequences,
suggesting that two populations with slightly different properties could occupy
these two very close loci.

\subsection{Towards a solution of the age spread problem}
\label{sec_ages}

The history of age spread determinations in $\omega$~Cen comprises a variety of
studies, all focused on the SGB, the most age sensitive region in the CMD. An
age spread of 2--5~Gyr, among the various sub-populations was first found by
\citet{hughes00} and \citet{hilker00}, based on the TO region morphology of
their high-quality Str\"omgren photometries and colour-metallicty calibrations.
Later several other papers came out based on combinations of high quality
photometry and low-resolution spectroscopy
\citep{hughes04,hilker04,rey04,sollima05b,stanford06,villanova07}, finding again
various age dispersions all around 2--5~Gyr. The only studies reporting on age
difference below 2~Gyrs are those by \citet{ferraro04}, who find that the SGB-a
cannot be fitted with any isochrone younger than the SGB-MP population, and
\citet{sollima05b}, who similarly found that the overall age spread of the SGB
sub-populations cannot amount to more than 2~Gyr. It is interesting to note that
\citet{sollima05b} and \citet{villanova07} use photometry from the same ACS data
set and spectra of similar quality, and reach opposite conclusions, with
\citet{villanova07} finding an age spread of at least 2~Gyr within the MP
population only. 

On the one hand, spectroscopic abundances of both RGB and SGB stars find an
$\alpha$-enhancement of all populations (except the RGB-a) consistent with pure
type II SNe enrichment which, in the self-enrichment scenario for $\omega$~Cen,
would imply very fast enrichmemt, within 1~Gyr. On the other hand, the high
$s$-process enhancement of all populations (except maybe the VMP found here)
would imply enrichment by intermediate mass (1--3~M$_{\odot}$) AGB stars on
longer timescales \citep[1--2~Gyr and possibly even more][]{busso99}. Therefore,
if $\omega$~Cen is a self-enriched system, we should find (and indeed many
authors do find) some significant age spread. While we cannot give the ultimate
solution to the relative ages (or age spread) puzzles listed above with the
present data, we can nevertheless use our newly defined VMP population to shed
some light on the problem.

We use isochrones from the Padova database of stellar evolutionary tracks and
isochrones\footnote{http://pleiadi.pd.astro.it/}, and in particular we chose the
ones based on the \citet{marigo08} and \citet{girardi00} tracks. We preferred
this set over BaSTI\footnote{www.oa-teramo.inaf.it/BASTI/}
\citep{pietrinferni06} simply because they provide tranformations to the WFI
filters. We note that the results in ages do not change significantly when using
BaSTI (less than 1$\simeq$Gyr overall shift), but the actual shape of the SGB is
better reproduced when the WFI filters are used instead. While the absolute ages
are somewhat uncertain, we show in Figure~\ref{fig_ages} that the age
differences are relatively robust regardless of the absolute age scale
calibration of the chosen isochones set. 

Firstly, we computed the Z values for the three populations using the formula
$$\log Z = \rm{[Fe/H]}-1.7+\log(0.638~10^{\rm{~[\alpha/Fe]}}+0.362)$$  \noindent
according to \citet{salaris93}. We adopted the \citet{lub02} reddening and
\citet{bellazzini04} distance modulus (see also Section~\ref{sec-param}). We
used the cosmological helium abundance models (for a discussion on higher He
abundances see Section~\ref{sec_he}). The uncertainties involved in the
isochrone fitting procedures are large: we assume aproximately $\pm$2~Gyr. We
find that the VMP and MP populations must be relatively close in age, with a
difference of 0$\pm$2~Gyr, while the MInt component should be about 2$\pm$2~Gyr
younger. If we force the age of the oldest population(s) to a cosmological value
of 13~Gyr, we have to increase E(B--V) and (m--M)$_V$ by 0.03 and 0.16~mag
respectively, and the fit becomes considerably worse when considering the
isochrones shape. The age differences, however, do not change. As a comparison,
we show in Figure~\ref{fig_ages} the case of NGC~6397, using the WFI B, V, and I
photometric catalogue described by \citet{carretta09} and the reddening and
distance modulus from the revised \citet{harris96} catalogue.

The existence of the VMP at the SGB level could help in solving two of the
paradoxes previously found in the literature. The first is the zero age
difference -- within the uncertainties -- found by \citet{sollima05b} between
the MP and the MInt populations, which is difficult to understand given the
difference in s-process enhancement between the two populations. If one assigns
[Fe/H]=--2~dex to the upper SGB envelope, and moves the MP population to branch
B, this shift in metallicity would be allowed by the uncertainties in the
\citet{sollima05b} low-resolution spectroscopy \citep[see also][on the
uncertainties of the calcium triplet calibration at low
metallicity]{starkenburg10}. As a result, the MP would become older, and the age
difference between the MP and MInt populations would not be so close to zero
anymore. 

The second puzzle that could be alleviated by the existence of the VMP is the
coexistence within the MP group of two separate populations with ages differing
by $\simeq$2--3~Gyr, found by \citet{villanova07}, which could be explained if
one admits that the MP population is in reality a mix of MP and VMP stars. The
metallicity difference between the VMP and MP populations would be difficult to
resolve with low-resolution spectroscopy, similarly to the case above. The
younger MP group identified by \citet{villanova07} would become older if its
metallicity was [Fe/H]$\simeq$--2.0~dex instead of --1.7~dex, thus recreating an
almost monotonic age-metallicity relation, although the age difference between
the VMP and the MP populations would be small. It could be argued that the
younger MP in \citet{villanova07} is much more numerous than the VMP estimated
fraction, i.e., 5\% of the total stellar content in $\omega$~Cen. However, we
must note that a large fraction of the targets in that study are selected {\em a
priori} on the upper SGB branch (or branch A), which we suspect is dominated by
VMP stars. In this case, the target selection would be biassed preferentially
towards the VMP population and the relative numbers in Figure~19 by
\citet{villanova07} would not be representative of the respective population
fractions anymore.

In a self-enrichment scenario, the small age difference between the VMP and the
MP population (0$\pm$2~Gyr) poses no problem as far as the type II SNe are
concerned, but could perhaps be too short to accommodate the $\simeq$0.5~dex
overabundance in s-process elements for the MP. On the other hand, if
2$\pm$2~Gyr occurred between the MP and the MInt2 population, there could be
enough time to enrich the MInt population in s-process elements up to the
observed level of [s/Fe]$\simeq$1~dex. Detailed chemical evolution calculations
would be extremely useful in understanding these details.

\subsection{The helium abundance}
\label{sec_he}

Since the discovery of a double MS in $\omega$~Cen \citep{anderson02,bedin04},
and the evidence that the bluer sequence was more metal-rich than the redder one
\citep{piotto05}, it was suggested that some of the metal-richer populations in
$\omega$~Cen could have abnormal He abundance \citep{norris04}, of about
Y$\simeq$0.35--0.40. A recent review by \citet{renzini08} discussed the possible
scenarios and compared them with observations of $\omega$~Cen, but also of
NGC~2808, which was found to possess a triple MS \citep{piotto07}, along with
other massive GGC.

Such an overabundant helium would have some impact in the model atmospheres
\citep{bohm79,girardi07} that are at the basis of any abundance analysis such as
the one presented here. The first thing to note is that the most metal-poor
populations of $\omega$~Cen do not require any He enhancement, and we assume
this to be the case not only for the two MP stars (WFI~503358 and 503951), but
also for the three VMP ones (WFI~507109, 507633, and 512938). The only star that
could suffer from an helium enriched atmoshpere is the one belonging to the
MInt2 population, WFI~512115. 

A simplified treatment by \citet{gray} assumes that an overabundance in helium
has a similar effect as an increase in gravity, as a first approximation:

$$\frac{\Delta g}{g} = \frac{4~\Delta A(\rm{He})}{1+4~A(\rm{He})}.$$

Just to derive an order-of-magnitude effect, we translate a mass fraction
increase from Y$\simeq$0.25 to Y$\simeq$0.35 into a number increase from
A(He)$\simeq$0.10 to A(He)$\simeq$0.15. This would be mimicked, in our
WFI~512115 star, by an increase in surface gravity from roughly 3.5 to 4.0~dex,
corresponding to an increase in abundance of 0.08~dex in [Fe/H]. Thus,
WFI~512115 would change from [Fe/H]=--1.19 to --1.11~dex, with basically no
significant effect in the age difference determination, within the quoted
uncertainties. We finally note here that, as discussed by \citet{sollima05b} in
their Figure~6, this increase in the helium abundance would only change the {\em
shape} of the isochrone's SGB, making it steeper, with a negligible (less than
1~Gyr) impact on the relative age determination.

\section{Summary and Conclusions}
\label{sec-concl}

We analysed UVES high-resolution spectra of six stars on the SGB of
$\omega$~Centauri. We compared our results with RGB high-resolution spectroscopy
to identify the sub-populations to which our targets belong, and we found
remarkable agreement with past abundance determinations. Three of our targets
(WFI~507109, 507633, and 512939) have [Fe/H]$\simeq$--2.0~dex, are
$\alpha$-enhanced and show no significant s-process enhancement. Two of the
remaining targets (WFI~503358 and 503951) belong to the MP population, with
[Fe/H]$\simeq$--1.65~dex, $\alpha$-enhanced and with [s/Fe]$\simeq$+0.5~dex. The
last target, star WFI~512115, belongs to the MInt2 population, with
[Fe/H]=--1.19~dex, $\alpha$-ehanced and with s-process enhancement similar to
the MP targets, i.e., slightly lower than what is expected for MInt stars. 

Our main result (see Section~\ref{sec_first}) is that {\em there exists an
additional, metal-poor population (that we name VMP) at [Fe/H]$\simeq$--2.0~dex,
which has chemical properties that make it the ideal candidate (remnant of) the
primordial population of $\omega$~Cen.} The RGB star ROA~213 \citet{smith00},
star 85007 by \citet{villanova10}, and 25 red giants studied by
\citet{johnson10}, have similar chemical composition and could represent the
prototype VMP members along the RGB. In particular, the s-process enhancement of
the SGB-VMP population is not compatible with the RGB-MP stars, while it is
compatible with the quoted RGB-VMP stars. Our conclusion is also supported by
previous work on metallicity distributions or RGB and SGB stars and by the
exquisite photometry by \citet{bellini10}. We estimate that this VMP population
should comprise at most 5\% of the entire stellar content of $\omega$~Centauri,
at present. The presence or absence of light element anti-correlations in this
population would be a fundamental constraint to the nature of $\omega$~Cen,
because anti-correlations are generally exclusively found in globular clusters
and never in the field populations of galaxies. From the available literature
\citep[mainly][]{johnson09,johnson10,marino10} it appears that
(anti-)correlations could be reduced in extent, in VMP stars. Until the presence
of (anti)-correlations in VMP stars is excluded by larger data samples, it looks
more promising to interpret this as the primordial population of $\omega$ Cen
instead of the remnant field population of its putative parent galaxy. 

The high-precision abundance determinations obtained allowed us to try to shed
some light on the SGB morphology relation with the RGB spectroscopically
identified sub-populations. We conclude that {\em there appears to be no
one-to-one correspondence between the nicely combed substructures of the RGB and
SGB. In particular, the MP and MInt populations could either be mixed along the
(lower) A, B, and C branches, or be positioned in a not strictly monotonic
order, with MInt1 occupying branch A (lower) and MP branch B, just as an
example.} As already said, VMP stars should occupy (and possibly dominate) the
uppermost SGB branch, or branch A in the \citet{villanova07} terminology.

We also found that (see Section~\ref{sec_ages}) {\em the existence of the VMP
population could alleviate some of the problems found in previous determinations
of relative ages.} In particular, the small metallicity difference between the
VMP and MP populations could have escaped previous abundance analyses based on
low-resolution spectra. The puzzling result by \citet{villanova07} that the MP
population should contain two groups with different ages could indeed be 
explained by the metallicity difference between VMP and MP. Also, the (too)
small age difference found by \citet{sollima05b} between the MP and MInt
populations woul become slightly larger when taking into account the existence
of the VMP, which should dominate the uppermost SGB envelope. 

Finally, {\em there should be a small age difference between the VMP and MP
populations (0$\pm$2~Gyr), while a slightly larger age difference (2$\pm$2~Gyr)
should occur between the VMP and the MInt2 populations.} Althought this latter
result is less secure because it relies on one star only, it agrees very well
with the majority of past studies \citep[see, e.g.,][]{stanford06}. The use of
different sets of isochrones (Padova, BaSTI) does not change the result
significantly, and the helium abundance problem should have a negligible impact
on the MInt2 star WFI~512115 (the only one which should have higher helium)
because, to first approximation, it should change its [Fe/H] by
$\simeq$0.08~dex. The age distribution suggested by the present data would
accommodate a fast enrichment between the VMP and MP populations, dominated by
SNe type II, while the s-process enrichment of the MP ([s/Fe]$\simeq$+0.5~dex)
could still pose a problem. The age difference between the MP and MInt
populations could instead be sufficient to allow for some intermediate-mass AGB
star enrichment \citep{busso99}, bringing [s/Fe] to +1.0~dex.

We conclude by noting that this is the only high-resolution-based abundance
analysis published on SGB stars in $\omega$~Cen so far. Even if the precision of
the abundances is higher than in past low-resolution studies, of course the
number of stars examined is only six. To give the final answer to the relative
ages problem in $\omega$~Cen, and to identify  who is who in the CMD at the SGB
level, a much larger sample (a few hundreds) of relatively high-resolution
spectra in the SGB region is absolutely necessary.

\begin{acknowledgements} 

We would like to thank A.~Sollima for help with isochrone fits, P.~Bonifacio,
and S.~Villanova for detailed comments on the abundance analysis and age spread
discussion. EP acknowledges the ESO {\em Visitor Programme} in Chile, where part
of this work was done.

\end{acknowledgements}

\end{document}